\documentclass[runningheads]{llncs}
\usepackage[T1]{fontenc}
\usepackage{graphicx}
\usepackage{multirow}
\usepackage{color}
\usepackage{colortbl}
\usepackage{booktabs}
\usepackage{xcolor}
\usepackage{tabularx}
\usepackage{underscore}

\usepackage[acronym]{glossaries}

\makeglossaries
\newacronym{IR}{IR}{Information Retrieval}

\newacronym{EE}{EE}{Evaluation Environment}
\newacronym{WT}{WT}{within time}
\newacronym{ST}{ST}{short-term}
\newacronym{LT}{LT}{long term}

\newacronym{ARP}{ARP}{Average Retrieval Performance}
\newacronym{nDCG}{nDCG}{Normalized Discounted Cumulative Gain}
\newacronym{ER}{ER}{Effect Ratio}
\newacronym{RI}{RI}{Relative Improvements}
\newacronym{DRI}{$\Delta \mathrm{RI}$}{Delta Relative Improvement}
\newacronym{RD}{$\mathcal{R}_e\Delta$}{Result Delta}

\newacronym{RRF}{RRF}{Reciprocal Rank Fusion}
\newacronym{d2q}{d2q}{Doc2Query}
\newacronym{LM}{LM}{Language Model}
\newacronym{ROS}{ROS}{ranking of systems}

\begin{document}
% General
\newacronym{IR}{IR}{Information Retrieval}

% Evaluation
\newacronym{EE}{EE}{Evaluation Environment}
\newacronym{WT}{WT}{within time}
\newacronym{ST}{ST}{short-term}
\newacronym{LT}{LT}{long term}

\newacronym{ARP}{ARP}{Average Retrieval Performance}
\newacronym{nDCG}{nDCG}{Normalized Discounted Cumulative Gain}
\newacronym{ER}{ER}{Effect Ratio}
\newacronym{RI}{RI}{Relative Improvements}
\newacronym{DRI}{$\Delta \mathrm{RI}$}{Delta Relative Improvement}
\newacronym{RD}{$\mathcal{R}_e\Delta$}{Result Delta}

% System
\newacronym{RRF}{RRF}{Reciprocal Rank Fusion}
\newacronym{d2q}{d2q}{Doc2Query}
\newacronym{LM}{LM}{Language Model}
\newacronym{ROS}{ROS}{ranking of systems}
\title{Replicability Measures for Longitudinal Information Retrieval Evaluation}
\author{Jüri Keller\inst{1}\orcidID{0000-0002-9392-8646} \and
Timo Breuer\inst{1}\orcidID{0000-0002-1765-2449} \and
Philipp Schaer\inst{1}\orcidID{0000-0002-8817-4632}}
\authorrunning{J. Keller et al.}
\institute{Technische Hochschule Köln, Ubierring 48, 50678 Cologne, Germany\\
\email{\{jueri.keller, timo.breuer, philipp.schaer\}@th-koeln.de}\\
\url{https://ir.web.th-koeln.de}}

\maketitle              % typeset the header of the contribution
\begin{abstract}
  Information Retrieval (IR) systems are exposed to constant changes in most components. Documents are created, updated, or deleted, the information needs are changing, and even relevance might not be static. While it is generally expected that the IR systems retain a consistent utility for the users, test collection evaluations rely on a fixed experimental setup. Based on the LongEval shared task and test collection, this work explores how the effectiveness measured in evolving experiments can be assessed. Specifically, the persistency of effectiveness is investigated as a replicability task. It is observed how the effectiveness progressively deteriorates over time compared to the initial measurement. Employing adapted replicability measures provides further insight into the persistence of effectiveness. The ranking of systems varies across retrieval measures and time. In conclusion, it was found that the most effective systems are not necessarily the ones with the most persistent performance.

\end{abstract}

\section{Introduction}
The environment of a retrieval system changes constantly. Not only but especially web retrieval systems are exposed to this due to the dynamic nature of the web. Documents, i.e., websites, get created, updated, or deleted~\cite{bar-ilanCriteriaEvaluatingInformation2002,dumaisTemporalDynamicsInformation2010}. But besides the evolving collection, the other components underlay change as well, from the information needs~\cite{dumaisPuttingSearchersSearch2014} to the relevance of search results~\cite{clarkeNoveltyDiversityInformation2008,tikhonovStudyingPageLife2013}.
These changes raise questions about the generalizability, temporal validity, and the persistency of \gls{IR} system effectiveness evaluations.

The LongEval shared task~\cite{DBLP:conf/clef/AlkhalifaBBCDEASGGKLLMPPSMZ23}\footnote{\url{https://clef-longeval.github.io}} seeks to investigate the temporal persistence of retrieval systems in a longitudinal evaluation. It, therefore, provides a first-of-its-kind web retrieval collection with sub-collections from different points in time~\cite{deveaudLongEvalRetrievalFrenchEnglishDynamic2023}. These sub-collections resemble the \gls{EE} a retrieval system is exposed to and allow to investigate how temporal changes influence retrieval systems~\cite{saezEvaluationInformationRetrieval2021}. The overall goal of the LongEval lab is to examine the \emph{temporal persistence} of retrieval systems. While the influence of temporal changes on the retrieved results are undeniable, it is unclear how the changes in effectiveness should be valued. For example, an over time increasing effectiveness would yield reliably good results. In this case, the users may profit, but the effectiveness would still change and quickly become unknown. Therefore, we argued that from an evaluation point of view, it can be desirable to investigate temporal reliability as persistence. In this work, we investigate the temporal persistence as a replicability task. Oriented at the ACM definition of replicability\footnote{\url{https://www.acm.org/publications/policies/artifact-review-and-badging-current}}, the goal is to achieve the same measurements in a different experimental setup, in this case, at a proceeded point in time. We investigate the temporal persistence of five advanced retrieval systems as a replicability problem. The systems are not specifically adapted to changes in the LongEval dataset to validate the temporal reliability of system-oriented \gls{IR} evaluations following the Cranfield paradigm. To facilitate reproducibility we make the code publicly available on GitHub.\footnote{\url{https://github.com/irgroup/CLEF2023-LongEval-IRC}}

\section{Related Work}
The LongEval dataset~\cite{alkhalifaLongEvalLongitudinalEvaluation2023} and shared task~\cite{DBLP:conf/clef/AlkhalifaBBCDEASGGKLLMPPSMZ23} provides the first test bed for investigating the temporal persistence of \gls{IR} systems. In the ongoing shared task, \gls{IR} systems are evaluated across three points in time and the relative change in effectiveness based on nDCG is measured by the \gls{RD}. Based on the submitted systems, no connection between effectiveness and temporal robustness was found but substantial correlation between the ranking of systems across the different points in time. 
Gon\-zález-Sáez~\cite{gonzalezsaez:tel-04547265} described different strategies for comparing IR systems in evolved environments. Beyond tracking one system across time also different systems are compared at different points in time. To maintain comparability different strategies are explored that use a pivot system, project scores to a common scale, or group topics into grains.

Directly related to the comparison strategy proposed in this work, Gon\-zález-Sáez et al.~\cite{saezEvaluationInformationRetrieval2021} achieve comparability by relating the results of different systems at different points in time to the same pivot system and compare only the measured deltas. In this work, also a pivot system is used but the same systems are compared in an environment with reduced dynamics. 

Besides the comparability of effectiveness, the temporal influences on test collections was investigated earlier by Soboroff~\cite{soboroffDynamicTestCollections2006}. He used the bpref measure to achieve a robuster ranking of systems on an evolving version of the GOV2 test collection. Further, he proposes indicators that describe how the test collection changed which can help to maintain it. Tonon et al.~\cite{tononPoolingbasedContinuousEvaluation2015b} describe test collection maintenance in an ``evaluation as a service'' methodology. To achieve reliable evaluations it is quantified how fair the current state of a test collection assesses a new system and estimates the cost of updating the test collection.

More works directly describe the changes in datasets, focusing on different components and granularity's~\cite{friederRepeatableEvaluationInformation2006,hopfgartnerContinuousEvaluationLargescale2019,dumaisPuttingSearchersSearch2014,clarkeNoveltyDiversityInformation2008,tikhonovStudyingPageLife2013}. These works are valuable sources to relate the changes in effectiveness back to the changes in the \gls{EE}.

\section{Temporal Replicability}
To analyze how the effectiveness evolves over time, we cast the longitudinal evaluation into a replicability task, i.e., we evaluate the same set of systems on different data. Naturally, a direct comparison of the measured effectiveness scores of the different \gls{EE}s is difficult since the recall base is not the same anymore. This makes it difficult to directly compare scores, and it remains unclear if the observed effects should be attributed to the system or the changing \gls{EE}. An advanced comparison strategy is necessary to overcome this problem~\cite{gonzalezsaez:tel-04547265}. In this work, we further explore the pivot strategy~\cite{breuerHowMeasureReproducibility2020a,saezEvaluationInformationRetrieval2021} in which the results of one system in one \gls{EE} are related to a pivot system that is evaluated in the same \gls{EE}. The delta between the experimental and the pivot system is then compared to a delta between the same systems measured in an evolved \gls{EE}. To align the terminology, the pivot system is a baseline run, BM25 for simplicity in this example, and the advanced run is the experimental system investigated. The intuition behind this evaluation strategy is that since the pivot system is exposed to the same \gls{EE} as the experimental system, hence encountering the same difficulties, it represents a neutral reference point that makes the results more comparable.

% RD
In the LongEval shared task, the \gls{RD} is used to describe how the effectiveness of the retrieval systems evolves over time. In this setting, the \gls{RD} is defined as reproduced and will serve as a baseline measure~\cite{DBLP:conf/clef/AlkhalifaBBCDEASGGKLLMPPSMZ23}:

\begin{equation}
  \mathcal{R}_e\Delta = \frac{\overline{M^{EE}(S)}-\overline{M^{EE'}(S)}}{\overline{M^{EE}(S)}}.
  \label{eq:RD}
\end{equation}

The \gls{RD} directly compares the mean retrieval effectiveness of a system $S$ quantified by a measure $M$ between the sub-collection \gls{EE} and \gls{EE}'. Improved effectiveness is denoted by a negative \gls{RD}, and values closer to 0 denote smaller changes which indicate more persistent systems.  

% DRI
In addition to the \gls{RD}, we adapt the \gls{DRI} and the \gls{ER}, initially proposed by Breuer et al.~\cite{breuerHowMeasureReproducibility2020a} as replicability measures, to investigate the temporal persistence of retrieval effectiveness. The replicability measures are implemented with the help of \texttt{repro\_eval}~\cite{breuerReproEvalPython2021a}, which is a dedicated reproducibility and replicability evaluation toolkit.

The \gls{DRI} describes how the effectiveness relatively changed from one \gls{EE} to an evolved \gls{EE}'. It is based on the \gls{RI} of an experimental system $S$ over the pivot system $P$. The \gls{RI} is adapted to the LongEval definitions as follows:

\begin{equation}
  \mathrm{RI} = \frac{\overline{M^{EE}(S)}-\overline{M^{EE}(P)}}{\overline{M^{EE}(P)}},  \qquad \mathrm{RI'}  = \frac{\overline{M^{EE'}(S)}-\overline{M^{EE'}(P)}}{\overline{M^{EE'}(P)}}.
\end{equation}

$M^{EE}$ denotes the effectiveness score of a measure $M$, e.g., nDCG, determined on the sub-collection \gls{EE} or \gls{EE}' respectively. The \gls{DRI} is then defined as:

\begin{equation}
    \Delta \mathrm{RI}= \mathrm{RI} - \mathrm{RI}'.
\end{equation}

Comparing different sub-collections is straightforward. The ideal \gls{DRI} of 0 is achieved if the \gls{RI} is the same between both sub-collections, indicating a system that performs robustly over time. The more \gls{DRI} deviates from 0, the less robust is the system, whereas negative scores indicate a more effective experimental system $S$ in the evaluation environment $EE'$, and higher scores correspond to a less effective experimental systems than in the evaluation environment $EE$.

% ER
While the \gls{DRI} describes the change in effectiveness, the \gls{ER} describes the persistence of the effectiveness. It is originally defined by the ratio between relative improvements of an advanced run over a baseline run. The relative improvements are based on the per-topic improvements, which are adapted for changing \gls{EE}s as follows:

\begin{equation}
  \Delta M^{EE}_j = M^{EE}_j (S) - M^{EE}_j (P)
\end{equation}

where $\Delta M^{EE}_j$ denotes the difference in terms of a measure $M$ between the pivot system $P$ and the experimental system $S$ for the $j$-th topic of the evaluation environment \gls{EE}. Correspondingly, $\Delta' M^{EE'}_j$ denotes the topic-wise improvement in the evaluation environment \gls{EE}'. The \gls{ER} is then defined as:

\begin{equation}
  \mathrm{ER} \big(\Delta'M^{EE'}, \Delta M^{EE}\big) = \frac{\overline{\Delta'M^{EE'}}}{\overline{\Delta M^{EE}}}= \frac{\frac{1}{n_{EE'}}\sum^{n_{EE'}}_{j=1}\Delta'M^{EE'}_j}{\frac{1}{n_{EE}}\sum^{n_{EE}}_{j=1}\Delta M^{EE}_j}.
\end{equation}

More specifically, the mean improvement per topic between the pivot and experimental system on one sub-collection \gls{EE} in comparison to the effect on the other sub-collection \gls{EE}' is measured. Thereby, the \gls{ER} is sensitive to the effect size. If the effect size is completely replicated in the second sub-collection, the \gls{ER} is 1, i.e., the retrieval system is robust. If the \gls{ER} is between 0 and 1, the effect is smaller, indicating a less robust system with performance drops. If the \gls{ER} is larger than 1, the effect is larger, indicating performance gains caused by the change of the \gls{EE}.

\section{Experimental Evaluation}
The proposed measures are tested in an experimental evaluation based on the LongEval test collection. The test collection is limited to the queries that are present in all sub-collections to reduce the dynamics and improve interpretability. Five retrieval systems and a BM25 baseline are compared, and the results for different effectiveness, persistency, and replicability measures are reported.

\subsection{LongEval Test Collection}
\begin{figure}[t]
  \centering
  \includegraphics[width=0.8\linewidth]{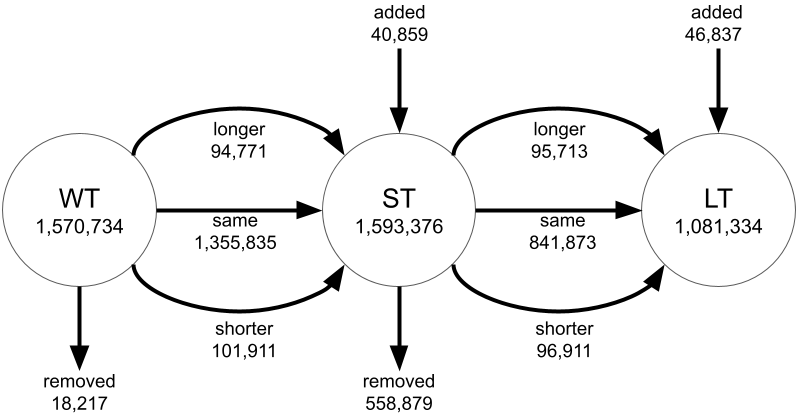}
  \caption{The evolution of the LongEval test collection documents across the three sub-collections. Over time, documents are added, removed, or updated. All documents were harmonized by their URLs.}\label{fig:longeval-doc-evolution}
\end{figure}

To our knowledge, the LongEval test collection~\cite{deveaudLongEvalRetrievalFrenchEnglishDynamic2023} is the first dataset specifically designed to investigate temporal changes in \gls{IR}. It consists of consecutive sub-collections that represent snapshots of a web search scenario evolving over time. The documents, topics, and qrels originate from the French, privacy-focused search engine Qwant.\footnote{\url{https://www.qwant.com/}} Logged user queries are selected as topics for the test collection, and the qrels are created from logged user interactions based on the Cascade Click Model~\cite{DBLP:conf/www/ChapelleZ09,DBLP:conf/wsdm/CraswellZTR08}. Therefore, the documents and queries are mostly in French, but there are also English machine translations available, which are mainly used in this work.
% Sub-Collections
The collections are organized into three sub-collections. The \gls{WT} sub-collection was created in June 2022. The \gls{ST} sub-collection was created in July~2022, immediately after the \gls{WT} collection. The third sub-collection, \gls{LT}, contains more distant data as it was created with a two-month gap from \gls{ST} in September 2022. 
% Documents
The changes in the document component are classified on a high level based on the string length in Fig.~\ref{fig:longeval-doc-evolution}. We note that between \gls{ST} and \gls{LT} considerably more documents are removed from the collections than between \gls{WT} and \gls{ST}.
% Topics
The topic sets also change across sub-collections, leaving a core set of 124 queries present in all sub-collections. The queries are typical keyword queries composed of at least one word and up to 11 words with few outliers. On average, a query consists of 2.5 words.
% Qrels
The qrels classify the documents' relevance on a three-graded scale, including \textit{not relevant}, \textit{relevant}, and \textit{highly relevant} labels. In general, the dataset has few assessed documents per topic. While the mean number of qrels is 14 per topic, the absolute number fluctuates between 2 and 59. Most of the documents are marked as not relevant, and the distribution of relevant and highly relevant qrels is skewed as well. Highly relevant documents are rare, with a maximum of only four and a mean of only one highly relevant document per topic. In the evaluations, these single documents heavily influence the final outcome as their position in the ranking especially impacts the score of rank-based measures like nDCG. For this work, we entirely rely on the English automatic translations of the test collection. 

\begin{figure}[t]
  \centering
  \includegraphics[width=\linewidth]{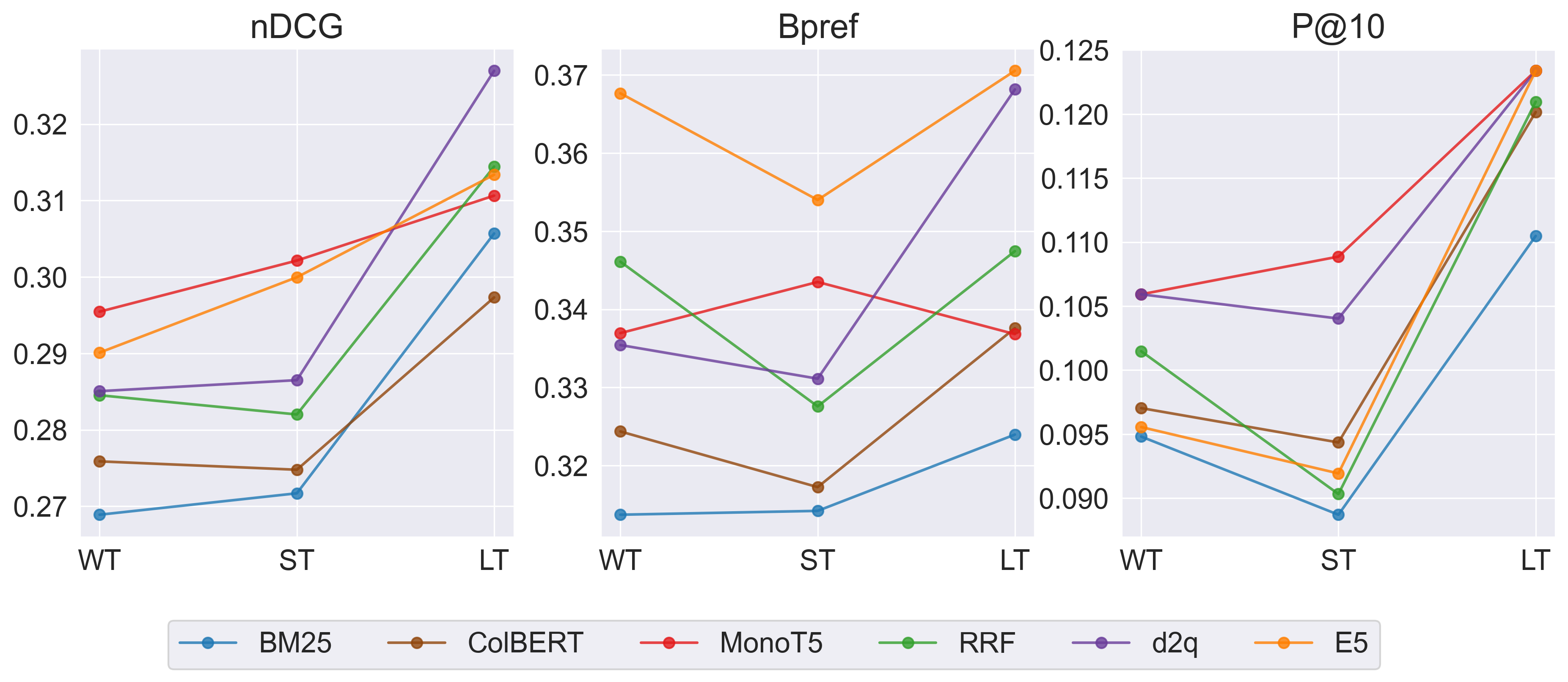}
  \caption{The P@10, bpref, and nDCG results based on the core queries.}
  \label{fig:ARP}
  \end{figure}

\subsection{Experimental Systems}
We compared different ranking functions and multi-stage retrieval systems on the \gls{WT} train slice of the LongEval dataset. The systems were selected as they represent state-of-the-art, off-the-shelf methods that are used in many recent \gls{IR} experiments. Therefore, it is especially interesting how these systems behave over time without being specifically adapted to a changing environment. The BM25~\cite{robertsonOkapiTREC31994} ranking function is used as the baseline and first-stage ranker for the advanced systems colBERT~\cite{khattabcolBERTEfficientEffective2020} and monoT5~\cite{pradeepExpandoMonoDuoDesignPattern2021}. Further, \gls{RRF}~\cite{cormackReciprocalRankFusion2009} of the runs from BM25 with Bo1~\cite{amatiProbabilityModelsInformation2003} reranking, DFR $\chi^2$ and PL2, E5\_base~\cite{wangTextEmbeddingsWeaklysupervised2022} as a dense retrieval system on the full dataset and d2q with ten expanded queries per document and BM25 as the retriever are tested. For a detailed description of the experimental systems, we refer the reader to the working notes~\cite{DBLP:conf/clef/Keller0S23} and the GitHub repository.\footnote{\url{https://github.com/irgroup/CLEF2023-LongEval-IRC}}

\subsection{Results}
For the evaluation of the result, the main goal is not a high but rather persistent performance. Therefore, the \gls{ARP} across EEs is compared to the \gls{RD}, and also the replicability measures \gls{DRI}, \gls{ER}, and the p-values of unpaired t-tests. The results measured by P@10, nDCG~\cite{DBLP:journals/tois/JarvelinK02}, and bpref~\cite{DBLP:conf/sigir/BuckleyV04} are reported in Tab.~\ref{tab:replicability}, and the \gls{ARP} is visualized in Fig.~\ref{fig:ARP}.

\begin{table}[t]
  \caption{Results of the persistency of effectiveness, measured on the core queries of the LongEval test collection. The replicability measures can not measure any persistancy for BM25 since this system is also used as the pivot. The ideal values of the replicability measures are noted at WT, the most persistent results are highlighted in bold, and results significantly different from BM25 at the same sub-collection are denoted by *.}
  \label{tab:replicability}
  \resizebox{\textwidth}{!}{
    \centering
    \begin{tabular}{cc|ccccc|ccccc|ccccc}
    \toprule
 & & \multicolumn{5}{c|}{P@10} & \multicolumn{5}{c|}{bpref} & \multicolumn{5}{c}{nDCG} \\
 & & ARP & $\mathcal{R}_e\Delta$ & $\Delta$RI & ER & p-val & ARP & $\mathcal{R}_e\Delta$ & $\Delta$RI & ER & p-val & ARP & $\mathcal{R}_e\Delta$ & $\Delta$RI & ER & p-val \\ \midrule

\multirow[c]{3}{*}{\rotatebox{90}{\scriptsize BM25}} 
& WT & 0.095 & 0      & -   & - & - & 0.314 & 0      & - & - & - & 0.269 & 0      & - & - & - \\
& ST & 0.089 & 0.064  & -   & - & - & 0.314 & \textbf{-0.002} & - & - & - & 0.272 & -0.010 & - & - & - \\
& LT & 0.110 & \textbf{-0.165} & -   & - & - & 0.324 & -0.033 & - & - & - & 0.306 & -0.137 & - & - & - \\\toprule

\multirow[c]{3}{*}{\rotatebox{90}{\scriptsize colBERT}} 
& WT & 0.097 & 0      & 0       & 1 & 1 & 0.324 & 0 & 0 & 1 & 1 & 0.276 & 0 & 0 & 1 & 1 \\
& ST & 0.094 & 0.028  & -0.040  & 2.540 & 0.858 & 0.317 & 0.022  & 0.024  & 0.286 & 0.826 & 0.275 & \textbf{0.004}  & 0.015 & 0.441  & \textbf{0.967} \\
& LT & 0.120 & -0.238 & -0.064  & 4.355 & 0.178 & 0.338 & -0.041 & -0.008 & 1.278 & 0.668 & 0.297 & -0.078 & 0.053 & -1.198 & 0.412 \\
 \toprule
\multirow[c]{3}{*}{\rotatebox{90}{\scriptsize monoT5}} 
& WT & \textbf{0.106} & 0 & 0 & 1 & 1 & 0.337 & 0 & 0 & 1 & 1 & \textbf{0.295} & 0 & 0 & 1 & 1 \\
& ST & \textbf{0.109} & -0.028 & -0.110 & \textbf{1.815} & 0.857 & 0.344 & -0.019 & -0.019 & 1.261 & 0.850 & \textbf{0.302} & -0.023 & -0.013 & 1.146 & 0.817 \\
& LT & \textbf{0.123} & \textbf{-0.165} & \textbf{0.000} & \textbf{1.161} & \textbf{0.332} & 0.337 & \textbf{0.000} & 0.034 & 0.553 & \textbf{0.997} & 0.311 & \textbf{-0.051} & 0.083 & 0.187 & \textbf{0.580} \\ \toprule

\multirow[c]{3}{*}{\rotatebox{90}{\scriptsize RRF}} 
& WT & 0.101 & 0 & 0 & 1 & 1 & 0.346* & 0 & 0 & 1 & 1 & 0.285* & 0 & 0 & 1 & 1 \\
& ST & 0.090 & 0.110 & 0.052 & 0.242 & 0.453 & 0.328 & 0.054 & 0.032 & 0.574 & 0.784 & 0.282 & 0.009 & \textbf{0.003} & \textbf{0.925} & 0.945 \\
& LT & 0.121 & -0.192 & -0.025 & 1.573 & 0.237 & 0.347* & -0.004 & \textbf{0.002} & \textbf{1.007} & 0.756 & 0.314 & -0.105 & 0.013 & \textbf{0.786} & 0.227 \\\toprule

\multirow[c]{3}{*}{\rotatebox{90}{\scriptsize d2q}} 
& WT & \textbf{0.106*} & 0 & 0 & 1 & 1 & 0.335 & 0 & 0 & 1 & 1 & 0.285 & 0 & 0 & 1 & 1 \\
& ST & 0.104* & \textbf{0.018} & -0.056 & 1.379 & \textbf{0.911} & 0.331 & 0.013 & \textbf{0.015} & \textbf{0.779} & \textbf{0.894} & 0.287 & -0.005 & 0.006 & 0.916 & 0.960 \\
& LT & \textbf{0.123} & \textbf{-0.165} & \textbf{0.000} & \textbf{1.161} & 0.326 & 0.368* & -0.098 & -0.067 & 2.034 & 0.300 & \textbf{0.327*} & -0.147 & \textbf{-0.010} & 1.317 & 0.150 \\ \toprule

\multirow[c]{3}{*}{\rotatebox{90}{\scriptsize E5}} 
& WT & 0.096 & 0 & 0 & 1 & 1 & \textbf{0.368*} & 0 & 0 & 1 & 1 & 0.290 & 0 & 0 & 1 & 1 \\
& ST & 0.092 & 0.038 & \textbf{-0.029} & 4.355 & 0.815 & \textbf{0.354} & 0.037 & 0.045 & 0.738 & 0.692 & 0.300 & -0.034 & -0.025 & 1.333 & 0.720 \\
& LT & \textbf{0.123} & -0.291 & -0.109 & 17.419 & 0.111 & \textbf{0.371} & -0.008 & 0.028 & 0.863 & 0.931 & 0.313 & -0.080 & 0.054 & 0.362 & 0.382 \\ \bottomrule
\end{tabular}

  }
\end{table}

% ARP
The effectiveness is similar for the systems but varies across EEs. Overall, the results of the tested systems improves in the long run with few exceptions, as measured by bpref. Mainly in the second EE (\gls{ST}), weaker results are achieved. Also, the ranking of systems varies across time and measure. In the first two EEs, monoT5 performs well, only outperformed by E5 as measured by bpref. In the last EE (\gls{LT}), the d2q, RRF, and E5 systems perform better than monoT5, except on P@10.

% RD
The \gls{RD} reflects the general upward trend in effectiveness indicated by decreasing negative values. While the \gls{RD} at \gls{ST} is negative for all systems except RRF and colBERT measured by nDCG, regarding bpref it is also positive for E5 and d2q. The more the \gls{RD} diverges from 0, the larger is the relative change and the less persistent the system performs. Regarding the different measures the \gls{RD} is instantiated with, no strong agreement for the most persistent system can be found in \gls{ST}. d2q, BM25, and ColBER achieve the most persistent results on P@10, bpref and nDCG. For the \gls{LT} EE monoT5 achieves the most persistent results on all measures, accompanied by BM25 and d2q in P@10.

\begin{figure}[t]
\centering
\includegraphics[width=0.4\linewidth]{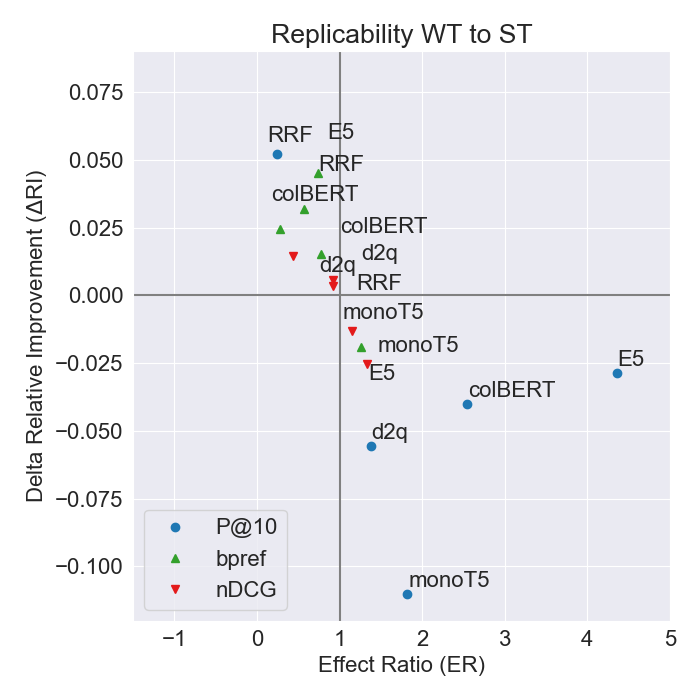}
\qquad
\includegraphics[width=0.4\linewidth]{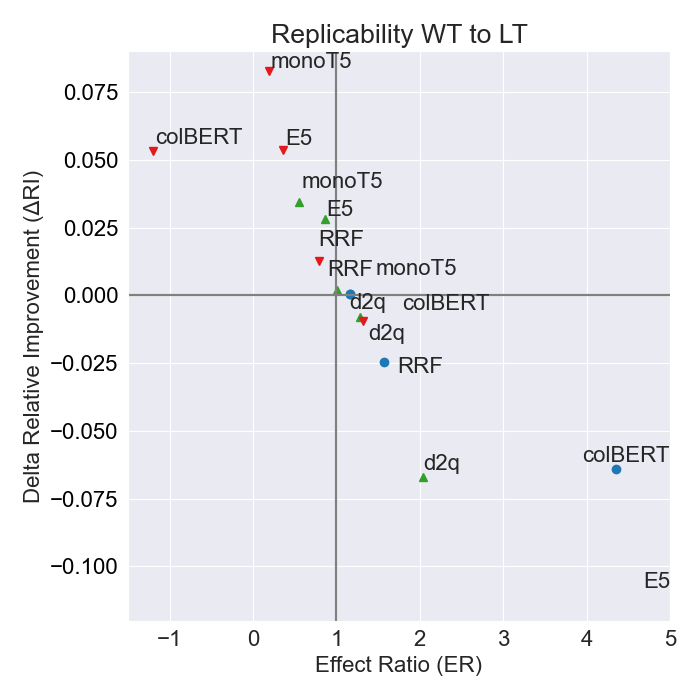}
\caption{The \gls{ER} plotted against the \gls{DRI} for the replication \gls{WT} to \gls{ST} (left) and \gls{WT} to \gls{LT} (right). The \gls{ER} for E5 is excluded as an outlier.}
\label{fig:DRI-ER}
\end{figure}

% Replicability
The \gls{DRI} and \gls{ER} complement the \gls{RD}. For instance, monoT5 achieved similar bpref scores on \gls{WT} and \gls{LT}, resulting in a \gls{RD} score of 0, which indicates perfect robustness in terms of \gls{RD}. However, when comparing \gls{DRI} and also \gls{ER}, more granular analysis is possible. In this case, the scores are close to but different from the perfect scores of 1 and 0, respectively, which would indicate perfect robustness. Regarding bpref, d2q achieves the best persistency according to \gls{DRI} and \gls{ER} in \gls{ST} and RRD in \gls{LT}. For the other measures, less agreement can be found.
% DRI - ER plot
The full potential of the \gls{ER} and \gls{DRI} can be seen if plotted against each other as in Fig.~\ref{fig:DRI-ER}. The closer the systems are located to the point (1, 0), the more persistent they are, with the preferable regions bottom right and top left. For the comparison of \gls{WT} to \gls{ST}, the monoT5 system performs well on bpref and nDCG. However, the effect and the absolute scores are slightly larger. E5 and monoT5 show large differences measured by P@10, with a larger effect for E5 (\gls{ER}) and a stronger improvement for monoT5 (\gls{DRI}). The RRF system, like most others, shows smaller absolute scores according to the \gls{DRI} and a slightly decreased \gls{ER}. The plot regarding \gls{WT} to \gls{LT} shows more outliers with larger effect sizes for P@10 for the E5 system (\gls{ER}=17.419) and bpref for the d2q system. The systems are shifted to the top right of the plot, a trend similar to the increased \gls{RD} for \gls{WT} to \gls{LT}.

\section{Discussion and Limitations}
% Persistancy
As initially mentioned, the notion of temporal persistence remains challenging to grasp. From the user's perspective, it might likely be desirable to always get the best results possible, even if the utility varies. Therefore, improving a system to perform more persistent is not beneficial, and direct implications for system design can not be derived. Instead, the potential in persistence evaluations lies in learning about the evaluation and test bed, quantifying the temporal validity of results, and the influence of the point in time when a test collection is created.

%%% Recall base changes Replicability
Comparing retrieval systems across time is difficult due to the changes in the experimental setups. It is unclear how to attribute the measured differences. Depending on the degree of change, a direct comparison of the ARP might not be sufficient or even meaningful since the recall base changes. As described before, in direct comparison, for example, through the \gls{RD}, the effect of the evolved environment is mainly extracted~\cite{saezEvaluationInformationRetrieval2021}. The replicability measures provide a method to abstract this effect to some extent and make the results comparable through the pivot system.
The experimental results showed that, in general, the \gls{RD} scores do not always agree on the most robust system with \gls{ER} and \gls{DRI}. Based on these findings, we conclude that the replicability measures provide another robustness perspective.
% Replicability agreement
We further see that it is not enough to consider the differences of a single retrieval measure like nDCG. Depending on the evaluation measure, different systems perform best in terms of robustness. For instance, \gls{RD} on \gls{ST} of nDCG is lower for colBERT and RRF than that of monoT5, while \gls{RD} of P@10 is lower or equal for monoT5. Similarly, the replicability measures should be instantiated with different retrieval measures to get a more comprehensive understanding of robustness. While the \gls{RRF} system achieves the best \gls{ER} instantiated with nDCG on both EEs, monoT5 is the most robust system in terms of \gls{ER} instantiated with P@10. Likewise, \gls{ER} and \gls{DRI} identify different systems as the most robust for the same measures and tasks, which shows that it is insightful to evaluate both replicability measures.

% p-val
In addition, we also included the p-values of unpaired tests based on the topic score distributions from different \gls{EE} that were determined with the same experimental system as proposed in~\cite{breuerHowMeasureReproducibility2020a}. The general idea of these evaluations proposes to assess the quality of replicability (in our case, robustness) by the p-values. It follows the assumption that lower p-values give a higher probability of failed replications or systems that are not robust. As can be seen, the highest p-values are achieved for the monoT5, colBERT, or d2q, which generally agrees with our earlier observations.

%%% Average per topic
\begin{figure}[t]
  \centering
    \includegraphics[width=\linewidth]{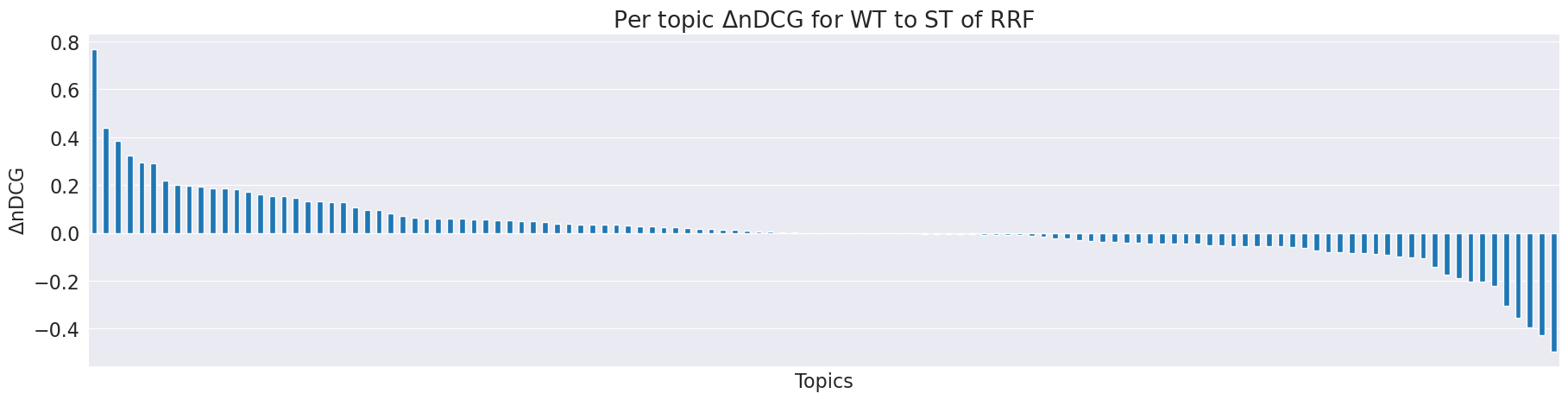}    
  \caption{RRF $\Delta nDCG$ results per topic for \gls{WT} to \gls{ST}. The topics are ordered according to the delta.}
  \label{fig:WTRRF}
\end{figure}

The \gls{RD} directly compares the results averaged across topics, but this \gls{ARP} may hide differences between the topic score distributions~\cite{breuerHowMeasureReproducibility2020a}. For example, the \gls{RRF} system achieved a high nDCG (0.285) at \gls{WT} and is relatively stable at \gls{ST} considering the \gls{RD} of 0.009. However, the per-topic results fluctuate between -0.4 and 0.8, as shown in Fig.~\ref{fig:WTRRF}. For some topics, the retrieval performance improves, while the changes in the \gls{EE} harm retrieval performance for other topics. We note that these circumstances require a more in-depth evaluation.

% Results are weaker
The experimental setup in this work limits the topic set to of the LongEval test collection to the core queries that are present in all sub-collections, thereby reducing the number of changing factors. In comparison, the effectiveness measured using the full test collection with all queries appears to be higher and demonstrates a stronger increase~\cite{DBLP:conf/clef/AlkhalifaBBCDEASGGKLLMPPSMZ23,DBLP:conf/clef/Keller0S23}. Generally, in this setting, the results for the different systems tend to be more similar. This is also reflected in the fewer significant differences per sub-collection between the experimental and the BM25 baseline system. Consequently, since only a few improvements are significant in this experiment, the ranking of systems is unreliable. While this may be negligible regarding the per-system comparisons across time, on which the replicability measures focus, it limits the general results. The fewer significant differences underscore the importance of the investigated retrieval scenario. Narrowing down the changes in the topics to those present in the core queries allows to attribute the measured effects to the changes in the document corpus, thereby improving interpretability. However, the measured effect also diminishes.

% What is the temporal connection contrebuting?
Further questions regard the relation between sub-collections. The disagreement between the \gls{RD} and the replicability measures might indicate the differences between sub-collections. While the sub-collections are related in time, it remains unclear what constitutes this context, especially regarding the effectiveness. This fosters the need to investigate what differentiates a longitudinal evaluation from a cross test collection evaluation.

%%% Limitations
This study is limited as it only considers the queries present in all sub-collections of LongEval, and no attempts were made to generalize across further test collections or retrieval scenarios. We note that the interpretation of results remains difficult, among others, because of the unintuitive notion of effectiveness persistence. Also, only BM25 was considered as pivot system for the replicability measures.

\section{Conclusion}
In this work, we investigated the utility of replicability measures to describe how persistent retrieval systems perform over time. We applied five retrieval systems to the LongEval test collection and quantified how the effectiveness changes.
The results showed that the retrieval effectiveness for most systems and measures increased over time on the LongEval dataset. The measured effectiveness deteriorates over time, which aligns with the natural assumption that results spanning longer timeframes are more different. Further, we report preliminary results applying replicability measures to quantify temporal persistence, an extension on common practices of these measures and their interpretation~\cite{DBLP:journals/ipm/MaistroBSF23}. It was shown that the results based on different measures and likewise for different topics do not necessarily agree with each other. Therefore, we see great potential in using replicability measures to gain further insights into robustness and also saw similarities to the measured result deltas. All in all, the strong influence of the experimental setup on the system's results could be shown and was analyzed. Since temporal persistence is a new challenge, interpreting the results is difficult. 

While these results are limited to the LongEval scenario, future work will extend the evaluation to further evaluation scenarios with different changes and dynamics~\cite{kellerEvaluationTemporalChange2024}. Aligning the documents of different sub-collections would enable to investigate the persistence on an even more specific level, for example, by casting the problem as a reproducibility task. Further open questions regard the selection of the pivot system to make the scores comparable and the selection of queries that allow meaningful temporal comparisons. Since the notion of temporal change remains difficult future work should regard generalizing persistence to temporal change. Lastly, an overall goal would be to employ the gained insights about temporal change to assess the temporal validity of evaluations.

\begin{credits}
\subsubsection{\ackname} We would like to express our gratitude to the LongEval Shared Task organizers for their invaluable efforts in constructing the LongEval dataset used in this study. Their dedication and hard work have provided an essential foundation for our research. We also gratefully acknowledge the support of the German Research Foundation (DFG) through project grant No. 407518790.

\subsubsection{\discintname}
The authors have no competing interests to declare that are
relevant to the content of this article.
\end{credits}

\bibliographystyle{splncs04}

\bibliography{bibliography-condensed}
\end{document}